\documentclass[12pt]{article}
\usepackage[dvips]{graphicx}
\newcommand{\eps}{\varepsilon}
\newcommand{\teta}{\vartheta}
\newcommand{\p}{\varphi}
\newcommand{\ctg}{\mathop{\rm ctg}\nolimits}

\newcommand{\w}[1]{\widetilde{\mathstrut #1}}

\author{ V.V.Sidorenko}
\title{On the dynamics of a spherically shaped top \\
                on a plane with friction }
\date{       Keldysh Institute of Applied Mathematics, \\
              Miusskaya sq. 4, Moscow 125047, RUSSIA    }
\begin{document}
\maketitle
\begin{abstract}
      The simple realistic model of  the  tippe  top  is
considered. An averaged system of equations of motion is obtained
in  special  evolutionary  variables.  Through  the   qualitative
analysis of this system the general features of the motion of the
top are obtained. Finally, some numerical results are presented.
\end{abstract}

\section*{1.Introduction}

     In this  paper  we  investigate  the  motion  of  a  top  of
spherical shape on a plane with sliding friction. Such a  problem
has been considered in  a  large  number  of  studies  concerning
the  ÿdynamics of the  tippe  top  -  a  children's  toy  with  the
paradoxical behavior ( the complete bibliography can be found  in
\cite{r1} - \cite{r7} ). Following O'Braien and Synge  \cite{r3},
we  postulate  a viscous friction law with the force of friction
proportional  to  the velocity of sliding and opposed to  that
velocity.  Although  this assumption has been  doubted  in  \cite{r6},
it  allows  us  to  explain and qualitatively predict the top's
possible behavior.

     The analysis will be restricted to motions which  are  close
to regular precessions. In order to write equations of motion  in
the form convenient for  application  of  asymptotic  methods  we
shall  use   the   special   evolutionary   variables   suggested
in \cite{r8,r9}.

\section*{2.Equations of motion}

     Let us consider  a  spherically  shaped  top  of  radius  $r$
placed on horizontal plane  $\Pi$ (ÿFig.1ÿ).  The  mass  distribution
inside the top is axially symmetrical. The mass center  $G$  of the
top lies at a distance $a$  from the  geometrical  center  of  the
top's surface. At the contact point $P$  the normal reaction force
${\bf N}$ÿ  and the force of friction  ${\bf F}$  are  applied  to
the  top.  As  mentioned in Sect.1, we  shall  assume  that  the
force  ${\bf F}$   is proportional to the velocity of the pointÿ$P$ÿ.

     To describe the equations of  motion  we  introduce  several
Cartesian coordinate systems. The system  $OXYZ$   is  a
spatially fixed coordinate system  with  the  axis  $OZ$
directed  upward; the  ÿplane $OXY$  coincides with  $\Pi$. The
coordinate  system $G\w{X}\w{Y}\w{Z}$ is originated in the top's
center of mass; the axes $G\w{X}$ ,  $G\w{Y}$,  $G\w{Z}$ are
parallel to to  the  axes  $OX$, $OY$, $OZ$   respectively.  The
coordinate system $G\xi\eta\zeta$  is fixed in a top's body;  the
axis   $G\zeta$  is directed along the symmetry axis. The fixed
coordinate  system orientation with respect to the system
$G\w{X}\w{Y}\w{Z}$ is defined by means of Euler's angles $\psi,
\teta, \p$ ( Fig.2 ). When   $\p = 0$ ,  the  fixed coordinate
system coincides with a semifixed system   $Gxyz$ . The axis
$Gx$  of the semifixed system is parallel to the  plane $\Pi$ ,
the axis  $Gz$  coincides with the axis  $G\zeta$.

\begin{figure}
\centering
\unitlength=1mm
\begin{picture}(0,85)
\put(-60,-75){\includegraphics[width=11.cm,height=15.0cm]{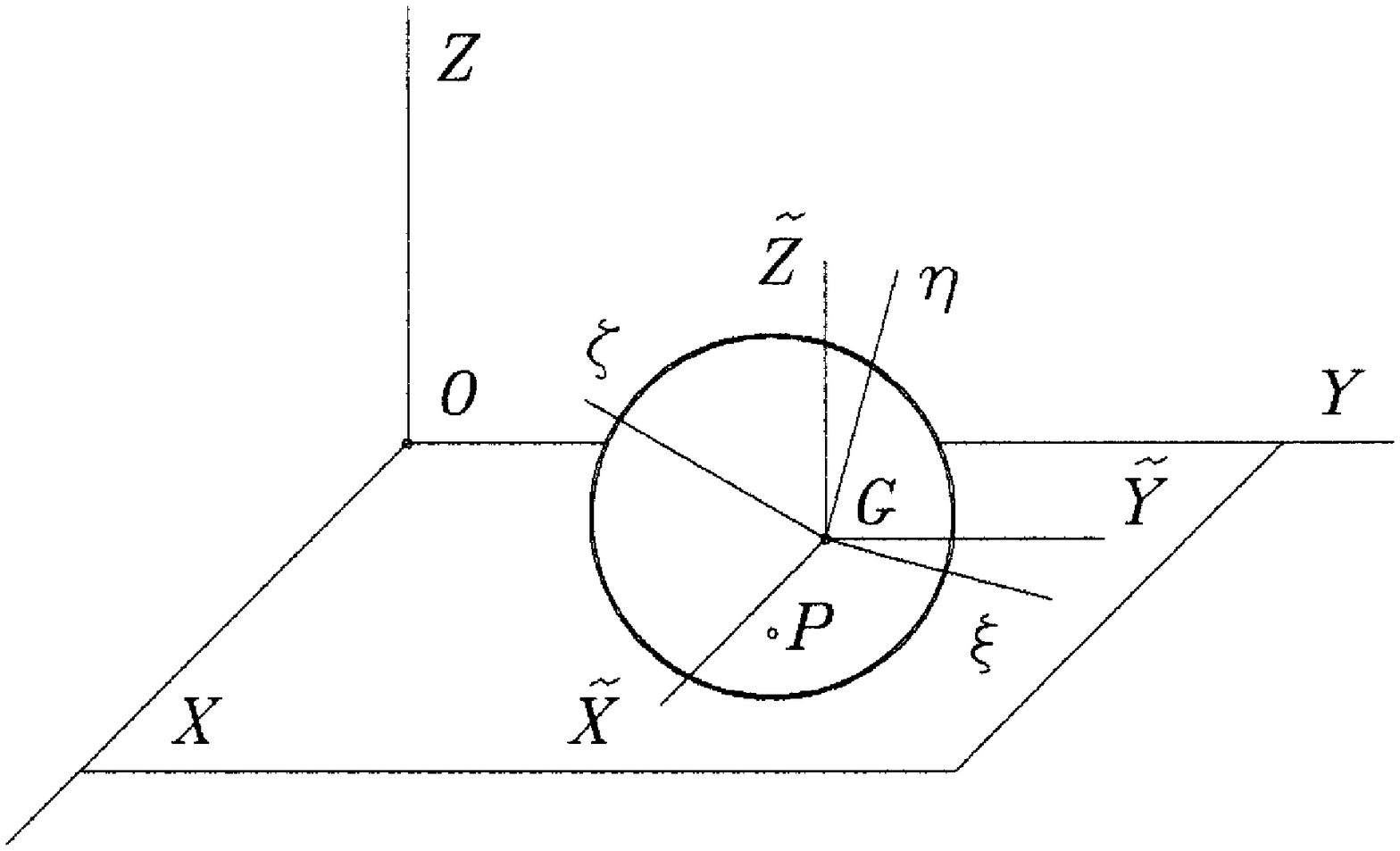}}
\end{picture}
\caption{A spherically shaped top on horizontal plane}
\end{figure}

     By using the Euler's angles  $\psi, \teta, \p$  and the
coordinates  $X_G  , Y_G$   of  the  top's center of mass in the
coordinate system $OXYZ$, we completely define the position of a
top on the plane. It should  be  note  that  in  case  under
consideration  when  the permanent contact of the  top  with the
plane  takes  place,  we have   $Z_G = r - a\cos\teta$.

     Dynamical  equations  for  the  top  on  the  plane  are   a
combination of the equations for the motion of  the  mass  center
and the equations for the motion of the top about its  center  of
mass. The motion of the mass center is described by the equations
$$
m{\dot V}_{GX}  = F_{X} \:  , \; m{\dot V}_{GY}  = F_{Y} \: .
\eqno (2.1a)
$$

\noindent
Here  $m$  is the mass of the top, $V_{GX}$  and  $V_{GY}$   are
the  components  of the mass center velocity in the system  $OXYZ$,
$F_X$  and  $F_Y$ are
the components of the force of friction in the same  system.  The
dots mean derivatives with respect to time  $t$  .  In  accordance
with the accepted assumption about the character of the  friction
we have
$$
F_X = - \eps |{\bf N}| V_{PX} \: , \; F_Y = - \eps |{\bf N}|
V_{PY} \: . \eqno (2.2)
$$

\noindent
where  $\eps$  is the coefficient of sliding friction,  $V_{PX}$    and
$V_{PY}$    are the components of the velocity of the contact point  $P$
in the system  $OXYZ$. The quantities  $V_{PX}$   and  $V_{PY}$    are
given by formulae
$$
V_{PX}  = V_{GX}  - \cos\psi [(r\cos\teta - a)\Omega_{y}
 + r\Omega_{z}\sin\teta] -  \sin\psi(r - a\cos\teta)\Omega_{x} \:
,
$$
$$
V_{PY}  = V_{GY}  - \sin\psi [(r\cos\teta - a)\Omega_{y}
 + r\Omega_{z}\sin\teta] +  \cos\psi(r - a\cos\teta)\Omega_{x} \:
. \eqno (2.3)
$$

\noindent
in which  $\Omega_{x}, \Omega_{y}, \Omega_{z}$  are the projections of
the  top's  angular velocity vector $\omega$ onto the axes  of  the  semifixed  coordinate
system  $Oxyz$.

\begin{figure}
\centering
\unitlength=1mm
\begin{picture}(0,70)
\put(-33,-30){\includegraphics[width=7.cm,height=10.0cm]{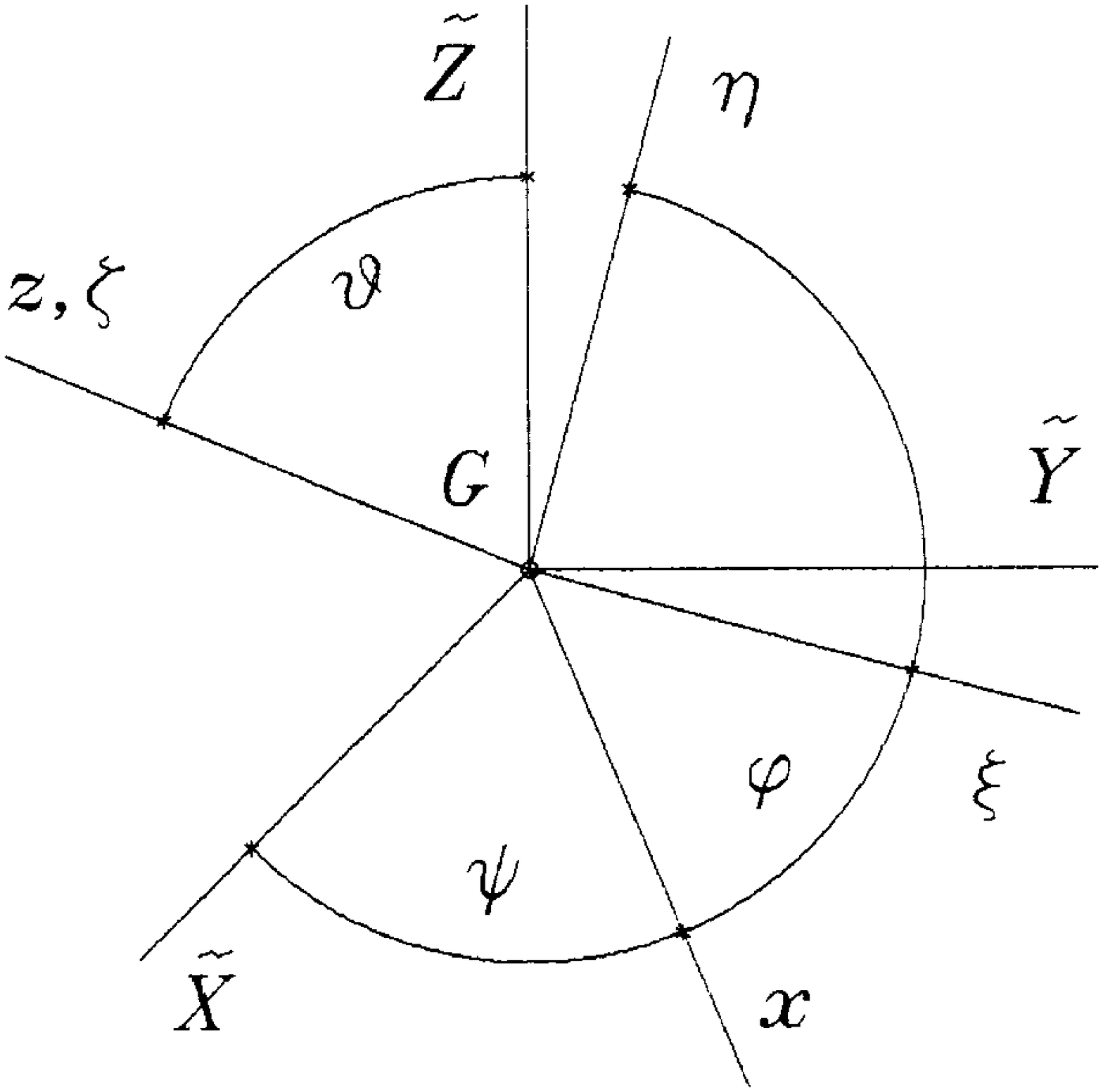}}
\end{picture}
\caption{The coordinate systems used}
\end{figure}

     The magnitude of  the  normal  reaction  force  ${\bf N}$   can  be
expressed as
$$
N = |{\bf N}| =  m[ g + a( \Omega^2_x \cos\teta + {\dot \Omega}_x
\sin\teta)]  .
$$

\noindent
Here  $g$  is constant of gravity.

     The equations for the motion of the top about the center  of
mass have the following form
$$
( A + ma^2 \sin^2 \teta) {\dot \Omega}_x  = -( C\Omega_z
 - A\Omega_y \ctg\teta)\Omega_y  +
\kappa\sin\teta[ 1 + \cos\teta (a\Omega^2_x/g)] + M_x \:,
$$
$$
A{\dot \Omega}_y  =
 ( C\Omega_z  - A\Omega_y \ctg\teta)\Omega_x  + M_y \:, \;
C{\dot \Omega}_z  = M_z \:,
$$
\smallskip
$$
\dot \psi  = \Omega_y / \sin \teta \: , \; \dot \teta  = \Omega_x
\: , \; \dot \p = \Omega_z - \Omega_y \ctg\teta \: . \eqno(2.1b)
$$

\noindent
Here  $C$  and  $A$  are axial  and  central  transverse  moments  of
inertia, $\kappa$  is equal to the product  $mga$ , $M_x , M_y , M_z$
are  the  projections of the torque caused by the friction force  onto  the
axes of the semifixed system. Taking into account  the  relations
(2.2),(2.3) we find
$$
M_x = - \eps (r - a\cos\teta)NV_{P*} \: ,
$$
$$
M_y = \eps (r\cos\teta - a)NV_{Px} \:, \; M_z = - \eps r
\sin\teta N V_{Px} \:,
$$
$$
V_{Px} =  V_{GX}\cos\psi + V_{GY}\sin\psi - (r -
a\cos\teta)\Omega_y  + r\Omega_z \sin\teta \:.
$$
$$
V_{P*} = -V_{GX}\sin\psi + V_{GY}\cos\psi + (r -
a\cos\teta)\Omega_x \:,
$$

     Dynamical equations (2.1) have Jellet's integral
$$
L = A\Omega_y \sin\teta + (r\cos\teta - a)C\Omega_z  = C_{\circ}
\: .
$$

     We shall suppose below that
$$
\eps \kappa^{3/2}A^{-1/2} \ll 1 \:, \;
|\omega| \sim {(\kappa/A)}^{1/2} \:,
$$
$$
{(V^2_{GX} + V^2_{GY})}^{1/2} \sim r(\kappa / A)^{1/2} \:.
\eqno (2.4)
$$
\smallskip

\noindent
When relations (2.4) take place, the influence of the friction on
the  motion  of  the  top  can  be  treated  as  a  certain  weak
perturbation.

     Under the proper choice of time, mass and length scales  the
sliding friction coefficient  $\eps$  will be  a  small  parameter  in
equations (2.1). It allows us to investigate the dynamics of  the
top on the plane by means of asymptotic methods.

\section*{3.Special evolutionary variables}

     First let us consider some  properties  of  the  unperturbed
motion. For  $\eps = 0$  system (2.1) describes the motion of
the  top along an absolutely smooth surface and has the first
integrals
$$
V_{GX} = C_1 \:, \;  V_{GY} = C_2 \:, \eqno (3.1a)
$$
\noindent
and
$$
u \equiv C\Omega_z = C_3 \: , \; w \equiv A\Omega_y \sin\teta +
C\Omega_z \cos\teta = C_4 \: , \eqno (3.1b)
$$
$$
E = \frac{1}{2} \left[m\left(V^2_{GX} + V^2_{GY}\right) + \left(A
+ ma^2 \sin^2\teta\right)\Omega^2_x  +  A\Omega^2_y
 + C\Omega^2_z\right] - \kappa \cos\teta = C_5
$$
\smallskip

Here  $u$  and  $w$  are the projections of the angular momentum onto
the axis of symmetry of  the  top  and  onto  the  vertical  axis
respectively, $E$  denotes the total energy of the top.

     In the  unperturbed  case  subsystem  (2.1b)  governing  the
rotational motion of the top reduces to the form
$$
( A + ma^2 \sin^2\teta){\dot \Omega}_x  = - ( C\Omega_z
 - A\Omega_y\ctg\teta)\Omega_y  +  \kappa
\sin\teta[ 1 + (a\Omega^2_x /g)] \:,
$$
$$
A{\dot \Omega}_y  =  ( C\Omega_z  - A\Omega_y\ctg\teta)\Omega_x
\:, \; C{\dot \Omega}_z  = 0 \: ,
$$
\smallskip
$$
\dot \p = \Omega_y /\sin\teta \:, \; \dot \teta = \Omega_x \:, \;
\dot \psi = \Omega_z  - \Omega_y \ctg\teta \: . \eqno (3.2)
$$
\smallskip

\noindent
Equations (3.2)  are  integrable  by  quadratures  \cite{r10,r11}.  In
general,  in  the  unperturbed  motion,  the  quantity  $\Omega_z$   is
constant, $\Omega_x, \Omega_y$  and $\teta$  are periodic functions  of
$t$   with  period  $T_{\teta}$ , $\psi$  and  $\p$  can be expressed
as follows

$$
\psi = \omega_{\psi}t + {\w{\psi}}(t) \: , \; \p
 = \omega_{\p}t   + {\w{\p}}(t) \:.
$$
\smallskip
\noindent Here  $\w\psi(t)$  and  $\w\p(t)$  are
$T_{\teta}$-periodic  functions of  $t$.  The frequencies
$\omega_{\teta} = 2\pi/T_\teta ,\, \omega_\psi$ and  $\omega_\p$
depend  in  a  complicated manner on the values of the first
integrals (3.1b) and in general are incommensurable.

     System  (3.2)  has  a  two-parameter  family  of  stationary
solutions
$$
\Omega_x \equiv 0 \: , \; \Omega_y \equiv \Omega_{y\circ} \: , \;
\Omega_z \equiv \Omega_{z\circ} \: , \; \teta    \equiv \Theta \:,
$$
$$
\psi = Wt + \psi_\circ  \:, \; \p = \omega_{\p\circ} + \p_\circ
\:. \eqno (3.3)
$$

\noindent
The constants  $\psi_\circ$  and  $\p_\circ$  in (3.3) are arbitrary,
while   $\Omega_{y\circ} , \Omega_{z\circ} , \omega_{\p\circ} , W$
and  $\Theta$  are connected by the relations

$$
\Omega_{y\circ} = W\sin\Theta \:, \; C\Omega_z = -
\frac{\kappa}{W} + AW\cos\Theta \:,
$$
$$
\omega_{\p\circ} = \frac{1}{C} \left[(A - C)W\cos\Theta -
\frac{\kappa}{W}\right] \:.
$$
\smallskip

\noindent
Solutions  (3.3)  correspond  to  those   motions  which  can  be
represented by a certain  superposition  of  a  uniform  rotation
about the axis of symmetry  and  a  uniform  rotation  about  the
vertical. Such motions are called ``regular  precessions''.  It  is
convenient to choose the velocity of the precession  $W$   and  the
angle of nutation  $\Theta$  as the parameters of the family (3.3).

     A closed subsystem of equations for $\Omega_x , \Omega_y$  and
$\teta$  can be derived from (3.3), containing  $\Omega_z$  as a
parameter. Setting

$$
\Omega_z  = \frac{1}{C} \left( - \frac{\kappa}{W} + AW\cos\Theta
\right) \: . \eqno  (3.4)
$$
\smallskip

\noindent we consider an  integral  manifold  $S_{W,\Theta}$
in  the  phase  space $( \Omega_x , \Omega_y , \teta )$ with a
fixed value for the integral $w$, pertaining  to the regular
precession with the parameters $W$  and $\Theta$ \cite{r12}. Its
parametric representation has the form

$$
S_{W,\Theta} =  \left\{ ( \Omega_x , \Omega_y , \teta ) \, : \:
\Omega_x = \Omega_x ( W, \Theta, c, \nu )\, , \: \Omega_y =
\Omega_y ( W, \Theta, c, \nu )\, , \right.
$$
$$
\left. \teta = \teta( W, \Theta, c, \nu )\: ; \:
0 \le \nu \le 2\pi \, , \:
0 \le c \le c_{\circ}(W,\Theta) \right\}
$$

\noindent where  $c$  and  $\nu$  denote the  amplitude  and the
phase  of  the nutational oscillations. At  individual solution
lying  on  the manifold  $S_{W,\Theta}$  $\nu = \omega_{\teta}t +
\nu_{\circ}$ . It  is not  difficult  to prove, through
Lyapunov's holomorfic integral theorem \cite{r13}, that the
functions  $\Omega_x ( W, \Theta, c, \nu )\, , \: \Omega_y ( W,
\Theta, c, \nu )\, , \: \teta( W, \Theta, c, \nu )$ can  be
written in the form of the series

$$
\Omega_x = \sum_{k=1}^{\infty}c^k \Omega_{xk}(W,\Theta,\nu) \: , \;
\Omega_y = \Omega_{y\circ} +
\sum_{k=1}^{\infty}c^k \Omega_{yk}(W,\Theta,\nu) \:,
$$
$$
\teta = \Theta + \sum_{k=1}^{\infty}c^k \teta_{k}(W,\Theta,\nu) \:.
\eqno (3.5)
$$
\smallskip

\noindent
which converge for sufficiently small values of $|\,c\,|$  (  to  apply
Lyapunov's theorem it is necessary to reduce  the  order  of  the
system for  $\Omega_x , \Omega_y , \teta$ using the integral  $w$  ).
We  have  the following expressions for the first coefficients

$$
\Omega_{x1} = \omega_{\circ}\sin\nu \:, \; \Omega_{y1} = -
(\kappa\cos\nu)/AW  \:, \; \teta_1  = \cos\nu  \:.
$$
\smallskip

\noindent Here $ \omega_{\circ} = \sqrt {(A^2 W^4  + 2\kappa A
W^2 \cos\Theta + \kappa^2)/  AW^2( A + ma^2 \sin^2 \Theta )} $ is
the fre\-quen\-cy of the small nutational oscillations.

     The  formulae  (3.4),(3.5)  define  the  local   change   of
variables

$$
(\, \Omega_x ,\, \Omega_y ,\, \Omega_z ,\, \teta \,) \;
\longrightarrow \;
(\, W ,\, \Theta , \,c ,\, \nu )
$$
\smallskip

\noindent
The new variables have a simple mechanical  meaning:  $W$   and   $\Theta$
specify the  reference  regular  precession,  while   $c$   and   $\nu$
characterize  the  amplitude  and   phase   of   the   nutational
oscillations  in  motion  which  is  close   to   the   reference
precession.  It  is  implied that this motion and  the  reference
precession   belong   to   the   same   joint   level   of    the
integrals  $u$  and  $w$.

     This change of  variables  reduces system (2.1) to  a  form
which  is  convenient  for  the  application  of  the   averaging
method  \cite{r14}.

     Variables  $W ,\, \Theta$  and   $c$   are  independent  integrals  of
unperturbed system. The following relations hold

$$
u = - \frac{\kappa}{W} + AW\cos\Theta \: , \;
w = - \frac{\kappa\cos\Theta}{W} + AW \: .
\eqno (3.6)
$$

\section*{4.Equations of motion of the top in the special \\
evolutionary variables }

     At first, we obtain equations for the variables   $W, \Theta$   by
means of two sequential substitutions:

$$
(\, \Omega_y ,\, \Omega_z\, ) \: \stackrel{1}{\longrightarrow} \:
(\, u ,\, w \,) \: \stackrel{2}{\longrightarrow} \: (\, W ,\, \Theta \,) \: .
$$
\smallskip

     For  $\eps \ne 0$  the change in the  projections  of  the  angular
momentum  onto  the  symmetry  axis  and  onto  the  vertical  is
described by the equations

$$
\dot u  = M_z \:, \;  \dot w  = M_z \cos\teta + M_y\sin\teta \: .
\eqno (4.1)
$$
\smallskip

\noindent
Expressing  $u$  and  $w$  in (4.1)  in  terms  of   $W$   and   $\Theta$  in
accordance with (3.6), we find

$$
\frac{\partial u}{\partial W}{\dot W} +
\frac{\partial u}{\partial \Theta}{\dot \Theta} =
- \eps r \sin\teta NV_{Px} \:,
$$
\smallskip
$$
\frac{\partial w}{\partial W}{\dot W} +
\frac{\partial w}{\partial \Theta}{\dot \Theta} =
- \eps a \sin\teta NV_{Px} \:.
\eqno (4.2)
$$
\smallskip

     Equations (4.2) define a  system  of  linear  equations  for
$\dot W$   and  $\dot \Theta$  with the de\-ter\-minant

$$
D = \frac{\partial(u,w)}{\partial(W,\Theta)} =
\omega^2_{\circ}\sin\Theta A(A + ma^2 \sin^2 \Theta)/W
$$
\smallskip

     System (4.2) can be solved if  $W \ne 0$  and  $\sin\Theta \ne 0$ :

$$
\dot W = - \eps\sin\teta NV_{Px} \frac{\partial L}{\partial
\Theta} \frac{\partial (W,\Theta)}{\partial (u,w)} \: ,
$$
\smallskip
$$
\dot \Theta = \eps\sin\teta NV_{Px} \frac{\partial L}{\partial W}
\frac{\partial (W,\Theta)}{\partial (u,w)} \: . \eqno (4.3)
$$
\smallskip

The substitution  $(\, \Omega_x, \,\teta\,)\: \longrightarrow (\,
c,\,  \nu\,)$ is analogous  to the Van  der  Pol
sub\-sti\-tu\-tion  \cite{r14}. Slightly modifying the Van   der
Pol approach, we find
$$
\dot c = \frac{\eps N}{\Delta_{\circ}} \left[ \frac{-(r -
a\cos\teta)V_{P*}}{A + ma^2\sin^2\teta} \frac{\partial
Q}{\partial \nu} + \sin\teta V_{Px} \left( \Delta^c_W
\frac{\partial L}{\partial \Theta} - \Delta^s_\Theta
\frac{\partial L}{\partial W} \right)
\frac{\partial(u,w)}{\partial(W,\Theta)} \right] \:,
$$
$$
\dot \nu = \omega_\teta - \frac{\eps N}{\Delta_{\circ}} \left[
\frac{-(r - a\cos\teta)V_{P*}}{A + ma^2\sin^2\teta}
\frac{\partial Q}{\partial c} + \sin\teta V_{Px} \left(
\Delta^{\nu}_W \frac{\partial L}{\partial \Theta} -
\Delta^{\nu}_\Theta \frac{\partial L}{\partial W} \right)
\frac{\partial(u,w)}{\partial(W,\Theta)} \right] \:,
$$
\smallskip

\noindent Here  $Q(W,\Theta,c,\nu) = \teta - \Theta$  and
functions $\Delta_{\circ} ,  \Delta^c_W , \Delta^c_{\Theta} ,
\Delta^{\nu}_W , \Delta^{\nu}_{\Theta}$    are defined by formulae

$$
\Delta_{\circ} =
\frac{\partial \omega_{\teta}}{\partial c}
\left(
\frac{\partial Q}{\partial \nu}
\right)^2 -
\omega_{\teta}
\left(
\frac{\partial Q}{\partial c} \frac{\partial^2 Q}{\partial \nu^2} -
\frac{\partial Q}{\partial \nu} \frac{\partial^2 Q}{\partial c \partial \nu}
\right) \:,
$$
\smallskip
$$
\Delta^c_W =
\frac{\partial \omega_{\teta}}{\partial W}
\left(
\frac{\partial Q}{\partial \nu}
\right)^2 -
\omega_{\teta}
\left(
\frac{\partial Q}{\partial W} \frac{\partial^2 Q}{\partial \nu^2} -
\frac{\partial Q}{\partial \nu} \frac{\partial^2 Q}{\partial W \partial \nu}
\right) \:,
$$
\smallskip
$$
\Delta^c_{\Theta} =
\frac{\partial \omega_{\teta}}{\partial \Theta}
\left(
\frac{\partial Q}{\partial \nu}
\right)^2 -
\omega_{\teta}
\left[
\left(
1 + \frac{\partial Q}{\partial \Theta}
\right)
\frac{\partial^2 Q}{\partial \nu^2} -
\frac{\partial Q}{\partial \nu} \frac{\partial^2 Q}{\partial\Theta\partial\nu}
\right] \:,
$$
\smallskip
$$
\Delta^{\nu}_W =
\frac{\partial Q}{\partial \nu}
\left(
\frac{\partial \omega_{\teta}}{\partial W}
\frac{\partial Q}{\partial c} -
\frac{\partial \omega_{\teta}}{\partial \nu}
\frac{\partial Q}{\partial W}
\right) -
\omega_{\teta}
\left(
\frac{\partial Q}{\partial W} \frac{\partial^2 Q}{\partial c \partial \nu} -
\frac{\partial Q}{\partial c} \frac{\partial^2 Q}{\partial W \partial \nu}
\right) \:,
$$
\smallskip
$$
\Delta^{\nu}_{\Theta} =
\frac{\partial Q}{\partial \nu}
\left(
\frac{\partial \omega_{\teta}}{\partial \Theta}
\frac{\partial Q}{\partial c} -
\frac{\partial \omega_{\teta}}{\partial \nu}
\frac{\partial Q}{\partial \Theta}
\right) -
$$
$$
\omega_{\teta}
\left(
\frac{\partial Q}{\partial \Theta} \frac{\partial^2 Q}{\partial c \partial \nu} -
\frac{\partial Q}{\partial c} \frac{\partial^2 Q}{\partial \Theta \partial \nu}
\right) -
\frac{\partial}{\partial c}
\left(
\omega_{\teta}
\frac{\partial Q}{\partial \nu}
\right) \:.
$$
\smallskip

     We will not write down awkward equations for  $\dot \psi , \dot \p ,
{\dot V}_{GX} , {\dot V}_{GY}$ in terms of special evolutionary  variables.
We  only  note that in general we have
$$
\dot \nu = O(1) \:, \; \dot \psi = O(1) \:, \; \dot \p = O(1)
$$
\noindent
while
$$
{\dot V}_{GX} = O(\eps) \, , \: {\dot V}_{GY} = O(\eps) \, , \:
\dot W = O(\eps) \, , \: \dot \Theta = O(\eps) \, , \: \dot c =
O(\eps c) \, .
$$
\noindent
Clearly, $V_{GX} , \, V_{GY} , \, W , \, \Theta , \,c$  are slow variables,
$\nu ,\, \psi ,\, \p$  are fast variables.

\section*{5.Averaged equations}

     We will analyze the behavior of  the  slow  variables  $V_{GX}, \,
V_{GY} , \, W , \, \Theta, \, c$  by the averaging method in the version
developed by
Volosov \cite{r15}. Because the right hand sides of the equations  of
motion do not depend on  the  proper   rotation   angle $\p$ ,  the
averaging along the unperturbed motion reduces to the independent
averaging with respect to $\nu$  and  $\psi$  ( in a nonresonant case  ).
The averaging procedure is complicated by the unevenness  of  the
variation of the fast variable $\psi$  due to  the  presence  of  the
periodic  component.  However,  taking  into  account   arguments
similar to those contained in \cite{r16}, this difficulty is easy  to
overcome.

     In the first approximation of the averaging method  we  find
( we  are  retaining  the  previous  notation  for  the  averaged
variables )

$$
{\dot V}_{GX} = - \eps g V_{GX}( 1 + O(c^2)) \: , \;
{\dot V}_{GY} = - \eps g V_{GY}( 1 + O(c^2)) \: ,
$$
\smallskip
$$
\dot W = - \eps m g \sin\Theta U
\frac{\partial L}{\partial \Theta}
\frac{\partial(W,\Theta)}{\partial(u,w)} + O(\eps c^2) \:,
$$
\smallskip
$$
\dot \Theta =  \eps m g \sin\Theta U
\frac{\partial L}{\partial W}
\frac{\partial(W,\Theta)}{\partial(u,w)} + O(\eps c^2) \:,
$$
\smallskip
$$
\dot c =
\eps m g c
\left[\Xi_1 + ( \Xi_2 + U\Xi_3 ) \frac{\partial(W,\Theta)}{\partial(u,w)}\right]
+ O(\eps c^2)
\eqno (5.1)
$$
\smallskip

\noindent Here  $U = \sin\Theta[r\Omega_{z\circ}(W,\Theta) - (
r\cos\Theta - a)W ]$   is  the  averaged projection of the
absolute velocity of the point  $P$  on  the   $Gx$ axis in the
regime of a  regular  precession  of  the  top  at  a velocity
$W$  with a nutation angle $\Theta$ , functions $\Xi_1(\Theta)
,\, \Xi_2(W,\Theta) ,\, \Xi_3(W,\Theta)$  are given by

$$
\Xi_1 = -  \frac{(r - a\cos\Theta)^2}{2(A + ma^2\sin^2\Theta)} \:
,
$$
\smallskip
$$
\Xi_2 =
\sin\Theta \left[ \frac{L}{2A} - (r - a\cos\Theta)W\right] \: ,
$$
\smallskip
$$
\Xi_3 =
\frac{\sin\Theta}{4\omega^2_{\circ}}
\frac{\partial(\omega^2_{\circ},L)}{\partial (W,\Theta)} -
[\cos\Theta - (ma^2\omega^2_{\circ}/2\kappa)\sin^2\Theta]
\frac{\partial L}{\partial W} \: .
$$
\smallskip

     As well as the original system (2.1), the averaged equations
have Jellet's integral

$$
L(W,\Theta) = rw - au = C_{\circ} \: .
$$

\section*{6.Qualitative analysis of the motion of the top \\
             on the basis of the averaged equations}

     The analysis of the averaged equations (5.1) reveals a  weak
interaction between rotational and horizontal motions of the top.
Indeed, in the equations for  ${\dot V}_{GX} , {\dot V}_{GY}$   the
influence  of  the rotational motion is expressed in negligibly small
terms of order  $\eps c^2$ . Omitting these terms, we find
$$
V_{GX}(t) = V_{GX}(0)e^{-\eps g t} \:, \; V_{GY}(t) =
V_{GY}(0)e^{-\eps g t} \:. \eqno (6.1)
$$

\noindent Thus  the  horizontal  component  of  the  center
mass  velocity asymptotically goes to  zero.  The
characteristic  time  of  the top's braking is  $T_{br}  = 1/\eps
g$ .

     Since the equations  for   $ {\dot W} , {\dot \Theta} , {\dot c}$
compose  a  closed subsystem, the evolution of the rotational
motion does not depend on the horizontal motion at all. In the
phase space $( W, \Theta, c )$ the condition  $c = 0$  defines
an  integral  manifold.  Solutions lying on the manifold tend
formally to regular  precessions  when $\eps \rightarrow 0$ .
These solutions have been investigated  in \cite{r1, r2, r7}.
Since there are no terms linear in  $c$  in the equations  for
${\dot W}$ and ${\dot \Theta}$ , small nutational oscillations
influence the behavior  of the variables $W$  and $\Theta$  in a
weak manner.

\begin{figure}
\centering
\unitlength=1mm
\begin{picture}(0,50)
\put(-70,-90){\includegraphics[width=14.cm,height=18.0cm]{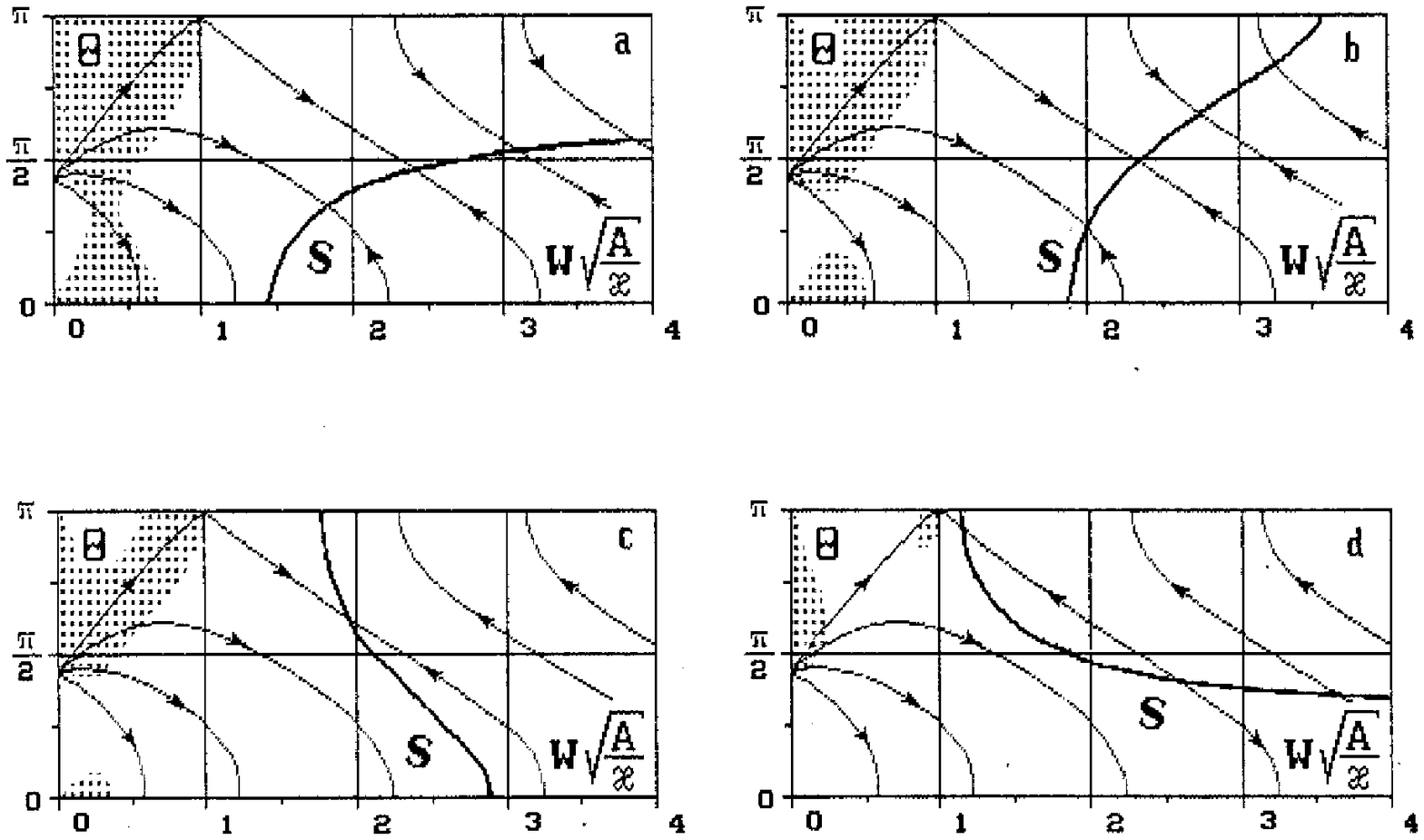}}
\end{picture}
\caption{Changes in the parameters of reference regular precession}
{(a -- $A/C=0.6666$, b -- $A/C=0.9$, c -- $A/C=1.1$, d -- $A/C=1.5$)}

\centering
\begin{picture}(0,90)
\put(-65,-70){\includegraphics[width=12.cm,height=15.0cm]{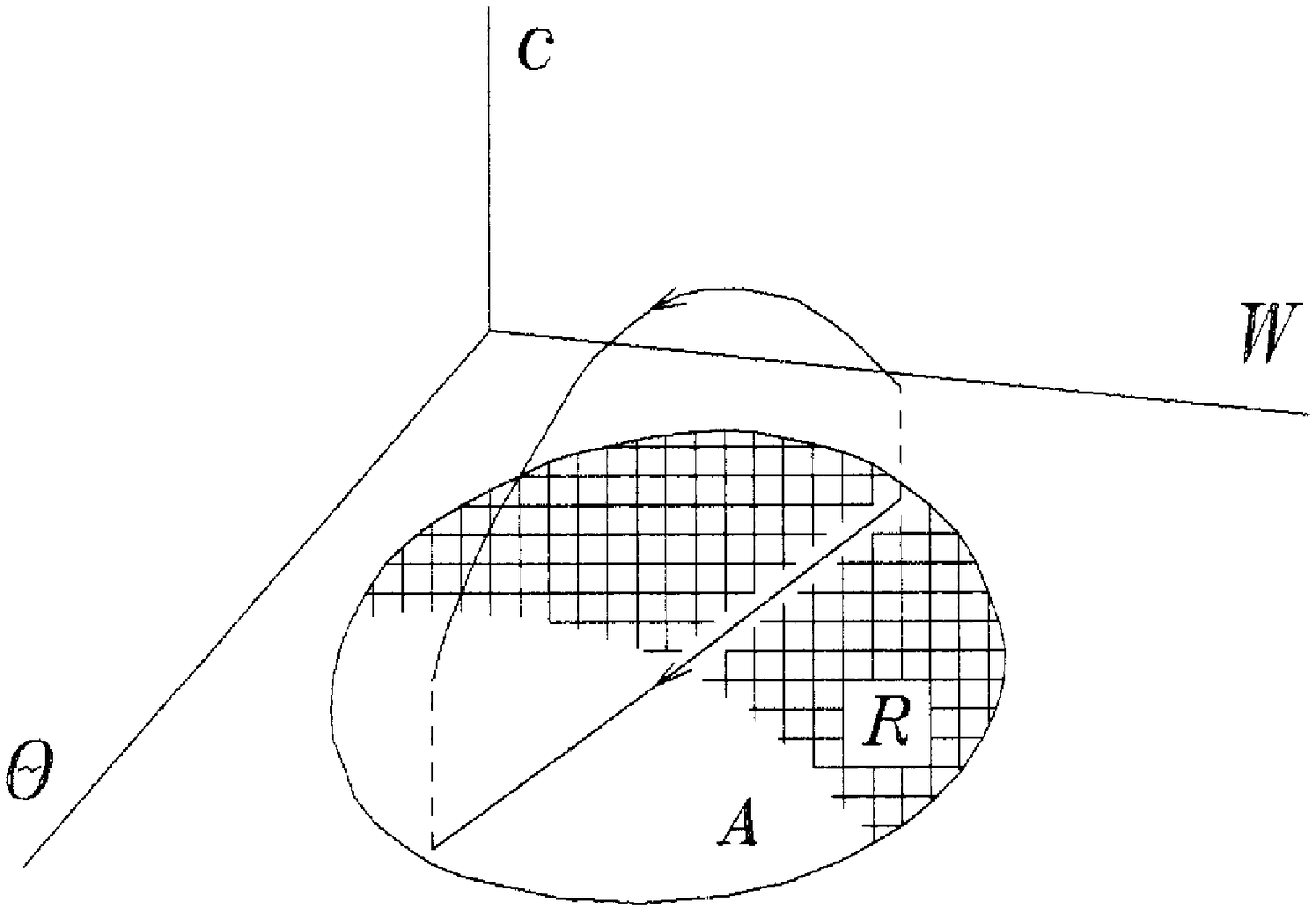}}
\end{picture}
\caption{Phase trajectory of averaged system.}
{$A$ and $R$ are attracting and repelling regions on the manifold $c=0$}
\end{figure}

     The phase portrait of the above-mentioned subsystem  on  the
manifold  $c = 0$   gives  us  a  clear  picture  of  the  general
properties of the rotational motion of the top.  As  an  example,
Fig.3 shows phase portraits constructed for a top with parameters
$r/a = 5, \,  ma^2 /A = 0.09$\,.  The  trajectories  are  the
levels   of Jellet's integral $L(W,\Theta) = C_{\circ}$.  Since
the  phase  portraits  are symmetric with respect to the axis $W
= 0$  , the  figures  can  be restricted to the regions of
positive values  of  the  precession velocity.

     Taking into account the sign of the expression in the square
brackets in the last equation (5.1), we can define the attracting
$([\cdots] < 0)$ and repelling $([\cdots] > 0)$ regions on the
integral manifold  $c = 0$ . On phase portraits the  repelling
regions  are shaded.  If  the   trajectory   drifts   above the
attracting ( repelling ) region near the manifold  $c = 0$ the
amplitude  of small nutational oscillations decreases ( increases
) ( Fig.4 ).

     Now let us consider stationary motions of  the  top  on  the
horizontal plane with friction. By setting in (5.1)
$$
{\dot V}_{GX} = {\dot V}_{GY} = {\dot W} = {\dot \Theta} = {\dot
c}  = 0
$$
\noindent
we can deduce the following relations:
$$
V_{GX} = V_{GY} = c = 0 \: , \;
\cos \Theta =
\frac{aCW^2 - r\kappa}{rW^2(C-A)} \: .
\eqno(6.2)
$$
\noindent
The relations (6.2) define a one-parameter family  of  stationary
motions ( curve  $S$  on the phase portraits in Fig.3 ).  In  these
motions  the  top  rolls  along   the   plane   without   sliding
$(\, V_{PX}=V_{PY}=0 \,)$ . Stability of the rolling without  sliding  has
been investigated in \cite{r17}.

     If geometrical and dynamical parameters of the  top  satisfy
the inequality
$$
aC > r| C - A | \eqno  (6.3)
$$
\noindent
the possible value  of  the  precession  velocity  in  stationary
motion is limited:

$$
\frac{r\kappa}{aC + r | C - A |}
\: < \,  W^2 \, < \,
\frac{r\kappa}{aC - r | C - A |}
$$
\smallskip

     In the case when the inequality  (6.3)  holds,  there  exist
solutions for which the angle of nutation increases from $\approx
\! 0$  to $\approx \! \pi$   ( Fig.3,b,c ).   These solutions
describe   the   top's overturning.

     When the velocity of the precession  is  large  enough,  the
equation for ${\dot \Theta}$  in (5.1) can be written in the form
$$
\dot \Theta =
\eps m g \sin \Theta
\left(r - a \cos \Theta\right)
\left[\frac{a}{A} - r\cos\Theta\left(\frac{1}{C} - \frac{1}{A}\right)\right] +
O\left(\frac{\eps}{W}\right) \: .
\eqno(6.4)
$$

\noindent
Using (6.4) it is easy to obtain that

$$
\arccos \left[ \tanh\left(C_{\Theta} - d_{-}t\right)\right]
\le \Theta(t) \le
\arccos \left[ \tanh\left(C_{\Theta} - d_{+}t\right)\right]
$$
\noindent
where
$$
C_{\Theta} = arcth\left(cos\Theta(0)\right) \: ,
$$
$$
d_{+} = \max_{0 \le \Theta \le \pi} \Gamma(\Theta) \le
\frac{1}{2}  \eps m g (r + a)
\left(\frac{a}{A} + r\left|\frac{1}{C} - \frac{1}{A}\right|\right) \:,
$$
$$
d_{-} = \min_{0 \le \Theta \le \pi} \Gamma(\Theta) \ge
\frac{1}{2}  \eps m g (r - a)
\left(\frac{a}{A} - r\left|\frac{1}{C} - \frac{1}{A}\right|\right) \:,
$$
$$
\Gamma(\Theta) = \frac{1}{2} \eps m g (r - a\cos\Theta)
\left[\frac{a}{A} + r\cos\Theta\left(\frac{1}{C} - \frac{1}{A}\right)\right]
$$
\smallskip

\noindent
The characteristic time for the top's overturning  is   estimated
by
$$
d^{-1}_{+} < T_{inv} < d^{-1}_{-} \: .
$$
     Finally we examine trivial stationary regimes in  which  the
top ro\-tates  uni\-formly  about  the  axis  of  dynamical  symmetry
directed  along  the  vertical.  The  trivial  regimes   can   be
represented by points at the upper and lower  boundaries  of  the
phase portrait. The behavior of trajectories near the  boundaries
are in the agreement with the results of  investigations  of  the
stability of trivial regimes \cite{r3}.

\begin{center}
\section*{7.Numerical simulation of the motion of the top}
\end{center}

     Properties  of  the  perturbed  motion   obtained   by   the
consideration of the  averaged  system  (5.1)  are  confirmed  by
results of numerical integration of the original system (2.1). In
addition, the numerical simulation reveals some new  features  of
the motion.

\begin{figure}
\centering
\unitlength=1mm
\begin{picture}(0,80)
\put(-70,-90){\includegraphics[width=14.cm,height=18.0cm]{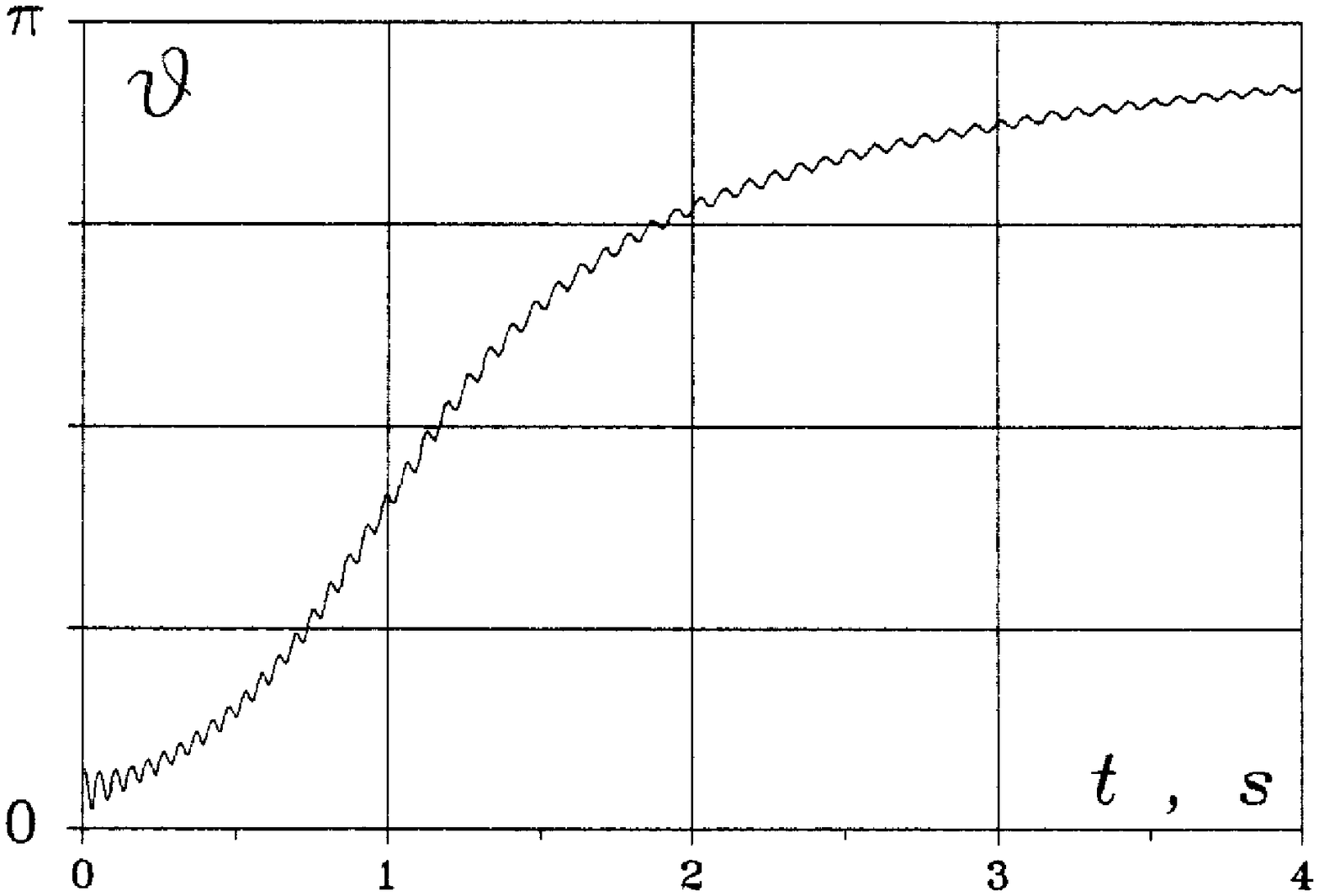}}
\end{picture}
\caption{Nutation angle of a rising top}
\end{figure}

     Fig.5 shows a typical plot of the nutation  angle  $\teta$  with
respect to the time  $t$  for the overturning top. We have chosen
$$
A = 10 \, g \!\cdot \! cm^2 \:, \; C = 9 \,  g \!\cdot \! cm^2
\:, \; m = 10 \,g \:,
$$
$$
a = 0.3 \, cm \:, \; r = 1.5 \, cm \:, \; \eps = 0.005 \, s/cm \:.
$$
\noindent
The initial values are
$$
V_{GX}(0)=V_{GY}=\Omega_x(0)=\psi(0)=\p(0)=0 \: ,
$$
$$
\Omega_y(0)=16.695 \, rad/s \: , \; \Omega_z(0)=129.372 \, rad/s
\: , \; \Theta(0)=0.241 \, rad \:.
$$
     The overturning process lasts  approximately  four  seconds.
The trajectory of the point $P$  on the  surface  of  the  top  is
displayed in Fig.6 ( we recall that  $P$   is  the  contact  point
between the top and the plane ). In the overturning process  this
point  moves  from  the  lower  hemisphere  to  the  upper   one.
Qualitatively the trajectory can be described as a  spiral  curve
which changes its direction in a vicinity  of  the  equator.  The
fine spikes and loops are caused by small nutational oscillations
of the top. It is not difficult to prove that  these  spikes  and
loops are oriented towards the nearest pole of  the  top  in  the
case $A > C$  and towards the the equator in the case  $A < C$ .
As an example the trajectory of the point $P$  on the  surface
of  a top with oblate ellipsoid of inertia is depicted in fig.7.
Here we have chosen
$$
A = 10 \, g \! \cdot \! cm^2 \:, \; C = 11 \, g \! \cdot \! cm^2
\:, \; m = 10 \,g \:,
$$
$$
a = 0.3 \, cm \:, \; r = 1.5 \, cm \:, \; \eps = 0.009 \, s/cm \:.
$$
\begin{figure}
\centering
\unitlength=1mm
\begin{picture}(0,80)
\put(-35,-40){\includegraphics[width=8.cm,height=12.0cm]{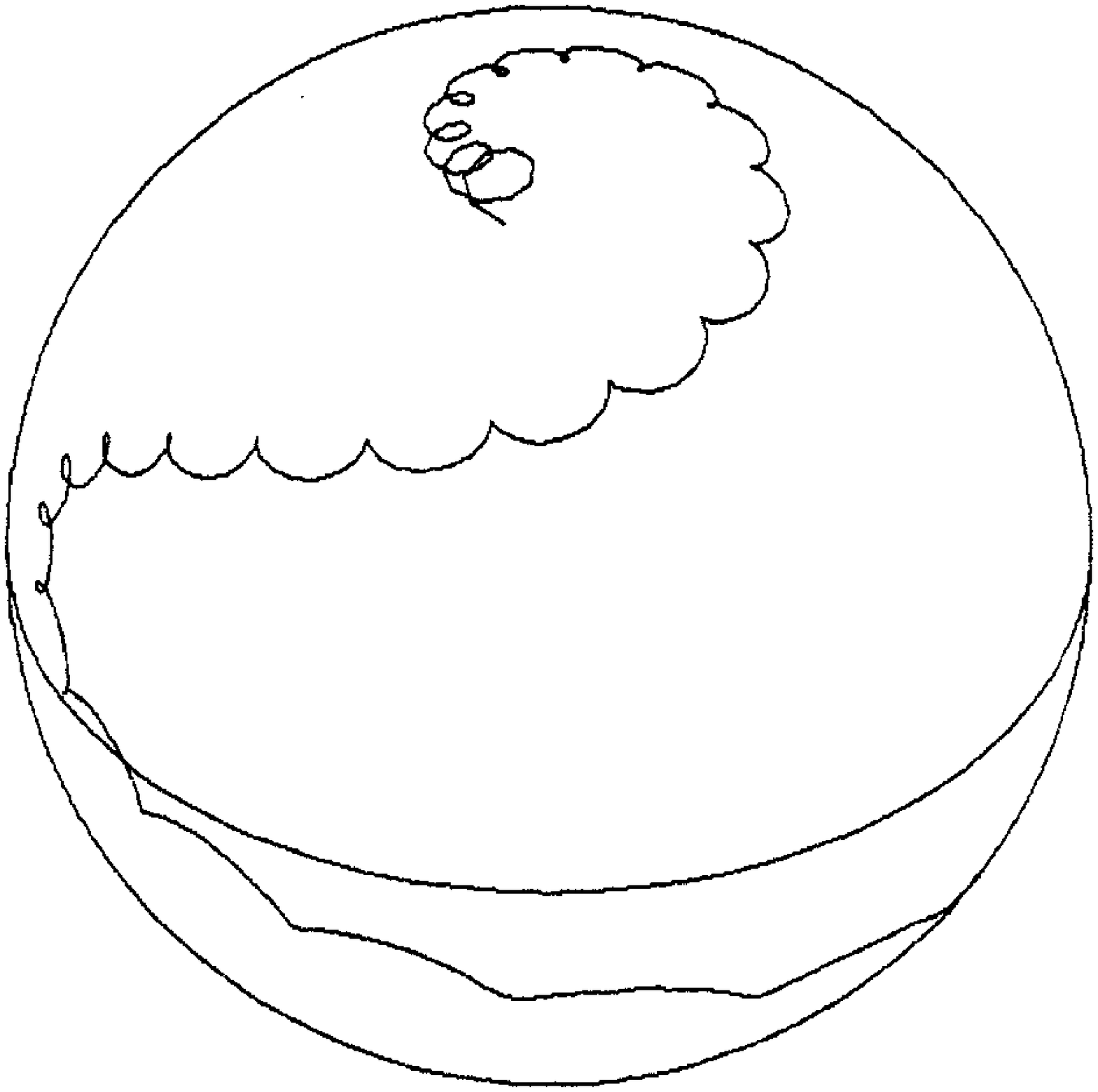}}
\end{picture}
\caption{Contact point trajectory in the case $A>C$}
{The top is overturned}

\centering
\begin{picture}(0,80)
\put(-35,-40){\includegraphics[width=8.cm,height=12.0cm]{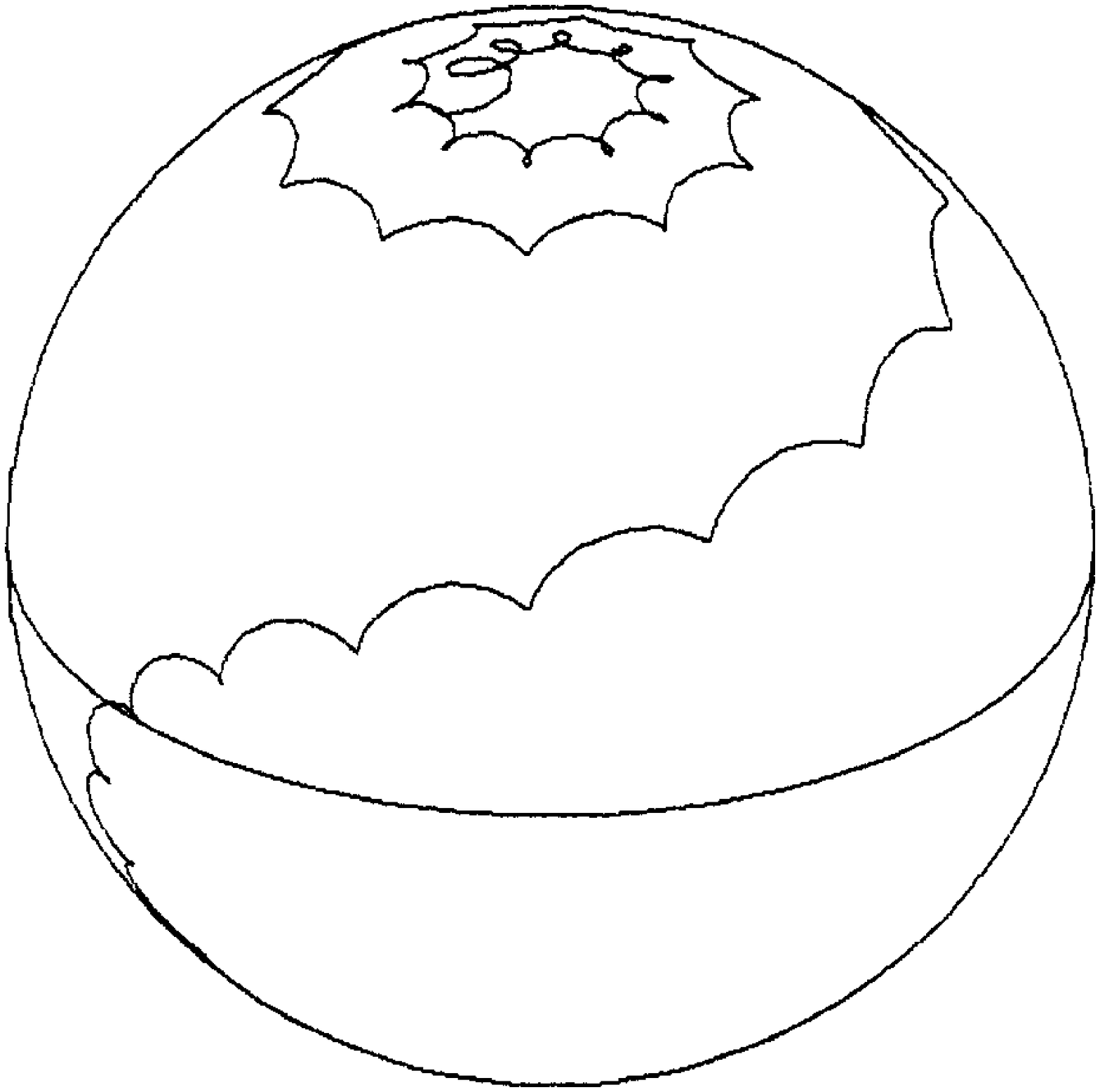}}
\end{picture}
\caption{Contact point trajectory in the case $C>A$}
{The top is overturned}
\end{figure}

\noindent
The initial values are
$$
V_{GX}(0)=V_{GY}=\Omega_x(0)=\psi(0)=\p(0)=0 \: ,
$$
$$
\Omega_y(0)=21.590 \, rad/s \: , \; \Omega_z(0)=137.244 \, rad/s
\: , \; \Theta(0)=0.241 \, rad \:.
$$
     It is interesting  to  compare  Fig.6  and  Fig.7  with  the
observed carbon traces after having spun an ordinary plastic  top
on the smoked glass \cite{r4}. Taking into account the orientation of
the trajectory loops in the photo of the top  in  \cite{r4},  we  can
conclude that this top has a prolate ellipsoid of inertia.

\section*{Conclusions}

     We present some new qualitative results on motion of the top
with spherical shape on a plane with friction.  The  results  are
obtained by perturbation analysis of differential equations which
describe the  dynamics  of  the  top.  The  special  evolutionary
variables were shown  to  be  convenient  for  carrying  out  the
averaging procedure.

\section*{Acknowledgments}

     The author is grateful for discussion with  G.G.Denisov  and
N.A.Fufaev.  For  helpful  comments  he  thanks  S.D.Furta.   The
research was partially supported by Russian Scientific Foundation
under Grant 93-013-16249.

\end{document}